\documentclass[9pt,conference]{IEEEtran}
\usepackage{blocked}
\usepackage{color}
\usepackage{multirow}
\usepackage{subcaption}
\usepackage {amsmath} 
\usepackage{stfloats} 
\usepackage{graphicx}
\usepackage[bookmarks=false]{hyperref}
\hypersetup{nolinks=true}
\usepackage{booktabs}
\usepackage{comment}
\begin{document}
%
\title{On the Dynamics of Solid, Liquid and Digital Gold Futures}
%
%

\author{\IEEEauthorblockN{1\textsuperscript{st} Toshiko Matsui}
\IEEEauthorblockA{\textit{Department of Computing} \\
\textit{Imperial College London}\\
London, SW7 2AZ, UK\\
t.matsui19@imperial.ac.uk}
\and
\IEEEauthorblockN{2\textsuperscript{nd} Ali Al-Ali}
\IEEEauthorblockA{\textit{Department of Physics} \\
\textit{Imperial College London}\\
London, SW7 2AZ, UK\\
ali.al-ali19@imperial.ac.uk}
\and
\IEEEauthorblockN{3\textsuperscript{rd} William J. Knottenbelt}
\IEEEauthorblockA{\textit{Department of Computing} \\
\textit{Imperial College London}\\
London, SW7 2AZ, UK\\
wjk@doc.ic.ac.uk}
}

%
%
\maketitle              
\begin{abstract}
This paper examines the determinants of the volatility of futures prices and basis for three commodities: gold, oil and bitcoin~-- often dubbed solid, liquid and digital gold~-- by using contract-by-contract analysis which has been previously applied to crude oil futures volatility investigations.
By extracting the spot and futures daily prices as well as the maturity, trading volume and open interest data for the three assets from 18th December 2017 to 30th November 2021, we find a positive and significant role for trading volume and a possible negative influence of open interest, when significant, in shaping the volatility in all three assets, supporting earlier findings in the context of oil futures. Additionally, we find maturity has a relatively positive significance for bitcoin and oil futures price volatility. Furthermore,
our analysis demonstrates that maturity affects the basis of bitcoin and gold positively~-- confirming the general theory that the basis converges to zero as maturity nears for bitcoin and gold~-- while oil is affected in both directions.


\end{abstract}
%

\section{Introduction}
\label{sec:introduction}




Recent increase in speculative trading activity within commodity futures, including Bitcoin futures, which is derived from the rise in financialisation of commodity markets, has affected the dynamics of these markets. These changes include the impact that open interest has in predicting commodity volatility \cite{Ripple2009} or returns \cite{Hong2012-futures}, and changes in the relation between futures prices across contract maturities \cite{Buyuksahin2008}.

Numerous studies have been undertaken on the determinants of futures price volatility. The earliest research in this domain, Samuelson \cite{samuelson1965proof}, demonstrated that futures prices become more volatile as contract maturity nears. However, it has since been shown that the trading activity of the futures contracts also affects the price volatility of futures contracts. Therefore, an investigation of not only the effect of maturity on futures commodities \cite{Anderson1985,Moosa2001_gold}, but also the role of trading volume and open interest is warranted. The former has been shown to positively affect futures volatility, while the latter provides a negative influence \cite{Bessembinder1993, HERBERT1995, Ripple2009}.


Despite the great deal of discussion on the determinants of futures volatility, little quantitative analysis has been carried out on the volatility of Bitcoin futures. Furthermore, there are few papers which attempt to explain basis using trading volume or open interest, in addition to maturity. Cryptocurrency research has developed since the genesis of Bitcoin \cite{SatoshiNakamoto2008}, however no consensus has been reached on its pricing efficiencies \cite{Urquhart2016,Mensi2019,CORBET201932} and the possible similarities with other assets \cite{CORBET201823}.

In light of these considerations, this study sets out to investigate the determinants of the futures price volatility for three commodities: Bitcoin, gold and oil. We use futures volatility as the objective variable, while taking maturity, trading volume and open interest as explanatory variables, with the intent of identifying the explanatory variables with the most predictive power for contract-by-contract data, as was previously applied to crude oil futures volatility analysis \cite{Ripple2009}. We then extend this methodology to examine the determinants of basis for the same three asset classes. The choice of gold and crude oil as assets to contrast with Bitcoin is due to similarities shared between the assets. Gold is chosen as Bitcoin is often considered to be an evolution of gold, especially as a store of value, being referred to as ``digital/new gold'' \cite{DYHRBERG2016139}. Crude oil is chosen as Bitcoin is also often referred to as a commodity \cite{BOURI2018224}. Moreover, Bitcoin shares common features with gold and crude oil, namely a well-defined circulating supply and a fixed total supply \cite{GRONWALD201986}. 

We compare several models with all possible combinations of objective variables in explaining the dependent variables, volatility and basis.
By employing contract-by-contract analysis for the three assets data from 18th December 2017 to date, we find futures volatility to be positively related to trading volume and negatively related to open interest in all three assets, supporting earlier results in the context of oil futures \cite{Ripple2009}. Furthermore, we demonstrate the positive and relatively significant role maturity plays for Bitcoin and oil futures price volatility. Our analysis also sheds light on role played by maturity in the determination of basis, affecting Bitcoin and gold futures positively -- supporting the general theory that the basis converges to zero at expiration for these two assets, but possibly not for oil.

\section{Preliminaries}

This section presents an overview of the key concepts of bitcoin futures, volatility and futures basis, relevant to this paper.

\subsection{Bitcoin futures}
The Chicago Board Options Exchange (CBOE) offered the first bitcoin futures on 10th December 2017, followed by the Chicago Mercantile Exchange (CME) on 18th December. Since the CBOE’s decision to delist bitcoin futures on 14th March 2019, the CME has been the largest bitcoin futures trading venue, with contract unit 5 bitcoin and tick size \$5 per bitcoin, yielding \$25 per contract \cite{CME-futures}.
On 3rd May 2021, the CME additionally started offering Micro Bitcoin futures, which are 1/10 the size of a standard bitcoin \cite{CMEmicro_press2021}, and options on Bitcoin futures. Furthermore, ProShares began trading of the first Bitcoin ETF on 19th October 2021, taking another step toward legitimising cryptoassets \cite{ETF_press2021}.

Widely considered to be a notable milestone in the evolution of an asset as an investment product, the launch of Bitcoin futures helped elevate Bitcoin's status to a station comparable to other financial assets \cite{AkyildirimETH2020}. 
The CME offers monthly contracts for cash settlement, meaning that an investor takes cash instead of bitcoin physical delivery upon settlement of the contract.

\subsection{Possible determinants of futures volatility}
\subsubsection{Time to maturity}
In terms of the major determinants of the volatility of futures prices, the concept of \textit{maturity effect} was firstly demonstrated by Samuelson \cite{samuelson1965proof}. The hypothesis states that the volatility of futures prices should increase closer to the contract maturity date, when the final payment is due on a financial instrument. This is based on the assumption that the futures market is more sensitive to the information regarding the underlying assets, the \textit{fundamentals}.

\subsubsection{Trading volume}
Trading volume is the number of completed trades in a specific security in a given time period. This is one of the important indices for measuring the trading activity of futures contracts. 

\subsubsection{Open interest}
\begin{figure}[h!]
    \centering
    \includegraphics[width = \columnwidth]{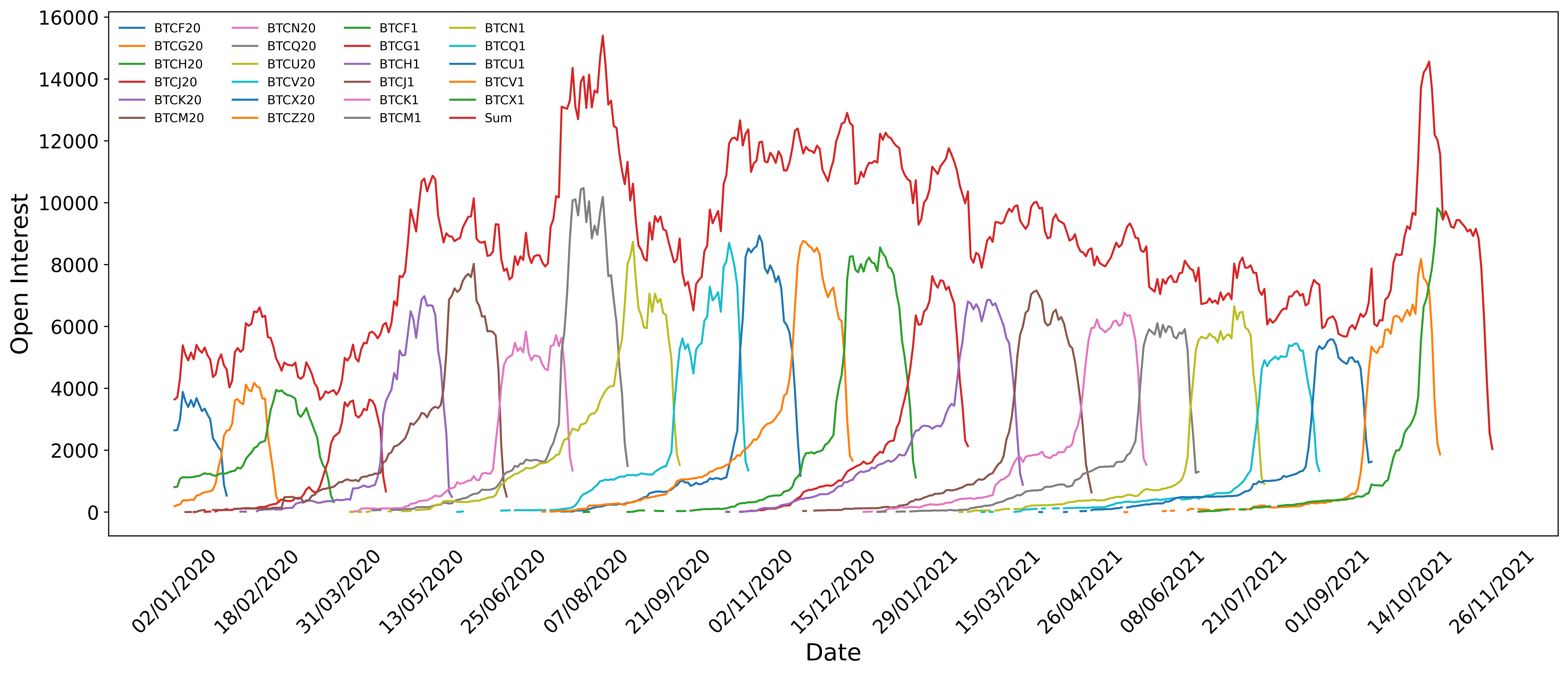}
    \caption{The evolution of open interest over the life of Bitcoin futures contracts and of the cumulative open interest across all active contracts from 1st January 2020 - 30th November 2021.}
    \label{fig:BTC_oi}
\end{figure}

Open interest refers to the total number of outstanding futures contracts that have not yet been settled.
It grows from zero when a contract is first listed, theoretically slumping back to zero upon contract maturity and typically peaking around the midpoint of the contract's lifetime \cite{Phan2020TTM}.
A rise in open interest coupled with a growth in price suggests an upward trend, while an increase or decrease in price while open interest remains flat or declines may implies a possible trend reversal \cite{murphy1999}. 
Fig. \ref{fig:BTC_oi} charts the evolution of bitcoin open interest for a subset of individual contracts, as well as the cumulative open interest across all active contracts. We can see that open interest approximately goes as a negative parabolic function of time-to-maturity for each contract. 

\subsection{Basis in the futures market}

In futures markets, \textit{basis} refers to the difference between the spot price of an asset and the corresponding futures price of that commodity. There have been a number of previous studies conducted on the basis since 1980s \cite{Modest1983-basis,KAWALLERIRAG1987-basis} and its relation to trading costs.

The basis is a critically important concept for portfolio managers, but the behaviour of the gap is not easily explainable until it converges to zero at the expiry of the nearest contract. The theoretical price of a futures contract at time $t$ is denoted as $F_{t} = S_{t}\mathrm{e}^{(r+u-y)*T}$,
where $S_{t}$ is the spot price of underlying asset, $r$ is the risk-free interest rate, $u$ is the storage cost per annum as a perecentage of the commodity price, $y$ denotes the convenience yield and $T$ is the time to maturity \cite{Hull_derivative}. This suggests an important role for maturity -- however, it is known that the basis mean-reverts, especially when the market is more liquid \cite{Roll2007-basis}.

Using this gap, traders constantly search for \textit{basis trade} opportunities, which involves buying a commodity at spot (taking a long position) and concurrently establishing a short position through derivatives.


\section{Data and Methodology}

We investigate the determinants of the futures price volatility of Bitcoin, gold and oil. Using futures volatility as the objective variable and maturity, trading volume and open interest as explanatory variables, we can reduce this to finding the significant explanatory variables for contract-by-contract data, as was previously applied to crude oil futures \cite{Ripple2009}. We then extend this methodology to evaluate the determinants of spot-futures basis for the same three asset classes. Based on the results, we aim to find the features unique to Bitcoin futures.


\subsection{Methodology}

As discussed in \cite{Ripple2009}, we employ a contract-by-contract analysis, which investigates individual contracts and decides the most meaningful set of data for a contract of each futures commodity. One empirical way to test this is to estimate the following regression:

\begin{equation}
dv_{i,c,t} = \beta_{0} + \beta_{1}m_{i,c,t} + \beta_{2}v_{i,c,t} + \beta_{3}o_{i,c,t} + \epsilon_{i,c,t}  \label{eq1-main}
\end{equation}

\noindent where $dv_{i,c,t}$ is either the futures volatility or basis on the day $t$ for contract $c$ of asset $i$. The \textit{volatility} is calculated as $(\ln(H)-\ln(L))^{2} / 4\ln2$, where $H$ and $L$ represent the high and low price for the day respectively. The \textit{basis} is calculated as the percentage difference between the spot and the futures prices, namely $(s_{t}-f_{t})/f_{t} *100)$. $m$ denotes the maturity, which is a decreasing counter, $v$ is the trading volume, $o$ represents open interest and $\epsilon$ is the residual regression term. For scaling purposes, $dv$ is multiplied by a factor of 10,000 and, only for gold and oil, $v$ and $o$ are measured in units of 10,000 contracts. The variables are not transformed into logarithms, following Hebert \cite{HERBERT1995} and Ripple \cite{Ripple2009}.


We first run Eq. (\ref{eq1-main}) with recent crude oil futures price data (i.e. $i =$ oil) to investigate the effects the three explanatory variables have on volatility, to back-test the Ripple's work \cite{Ripple2009}. We then extend the method to Bitcoin and gold to test whether there are any difference in the determinants of futures volatility between asset classes. Furthermore, we change the dependent variable, $dv$, to basis to explore if there is a similar trend in the determinants of basis, especially for Bitcoin.

\subsection{Data}\label{ssec:data}

The spot and futures contract series from 18th December 2017 to 30th November 2021 for Bitcoin, gold, crude oil were collected from \textit{Bloomberg}. We took daily data for spot price and futures series (price, trading volume and open interest) for each asset at closing time, resulting in 47 contracts, whose maturity dates are the last Friday of each month.
We employ the last two months of observations for each of the 47 contracts in order to include observations that reflect the focus of market activity \cite{SERLETIS1992, HERBERT1995, Ripple2009}.

\begin{table*}[bp] 
\centering
\setlength{\tabcolsep}{4pt}

\caption{Descriptive Statistics for Bitcoin, Gold and Oil Spot and Futures Markets}\label{tab:descriptive}
\begin{tabular}{lrrrrlrrrrlrrrr}
\toprule
    & \multicolumn{4}{l}{\bf Bitcoin}~~~ &~& \multicolumn{4}{l}{\bf Gold} &~& \multicolumn{4}{l}{\bf Crude Oil}\\
        \cline{2-5}
        \cline{7-10}
        \cline{12-15}
    \addlinespace[0.8ex]
    ~& Spot ~~~ & Futures ~ & ~ &&& Spot ~~~ & Futures ~ & ~&&& Spot ~~~ & Futures~ && \\
        \cline{3-5}
        \cline{8-10}
        \cline{13-15}
    \addlinespace[0.6ex]
    ~&  Price (\$)  & Price (\$)  & Volume~ & Open Int. ~&&  Price (\$)  & Price (\$)  & Volume~ & Open Int. ~&&  Price (\$)  & Price (\$)  & Volume~ & Open Int.  \\
\midrule
Mean & 17,432.9 & 17,547.3 & 3,407.5 & 2,580.6 & & 1,512.8 & 1,512.8 & 66,038.9 & 65,261.6 & & 56.528 & 56.613 & 380,442.0 & 310,568.9 \\
Median & 9,416.0 & 9,470.0 & 2,015.0 & 1,964.0 & & 1,474.6 & 1,474.6 & 752.0 & 1,967.0 & & 58.430 & 58.470 & 325,782.0 & 311,760.0 \\
Maximum & 67,734.0 & 67,905.0 & 31,433.0 & 10,475.0 & & 2,063.5 & 2,054.6 & 813,406.0 & 553,207.0 & & 83.870 & 83.870 & 228,823.0 & 642,781.0 \\
Minimum & 3,156.9 & 3,135.0 & 6.0 & 15.0 & & 1,174.2 & 1,176.2 & 1.0 & 1.0 & & -37.630 & 11.570 & 23,217.0 & 45.0\\
Std Dev. & 16,867.1 & 17,002.6 & 3,939.3 & 2,272.9 & & 244.52 & 244.47 & 128,498.6 & 125,421.3 & & 12.728 & 12.259 & 255,353.3 & 132,546.0 \\
Skewness & 1.44384 & 1.44575 & 1.86425 & 1.17051 & & 0.43934 & 0.43055 & 1.89599 & 1.82845 & & -1.09695 & -0.89745 & 0.95076 & -0.36228\\
Kurtosis & 3.61314 & 3.62582 & 7.71747 & 3.62054 & & 1.79948 & 1.77948 & 5.91323 & 5.04775 & & 5.07415 & 3.60436 & 4.62567 & 2.78014 \\

\bottomrule
\end{tabular}
\end{table*}

\begin{figure}[h]
\includegraphics[width=\columnwidth]{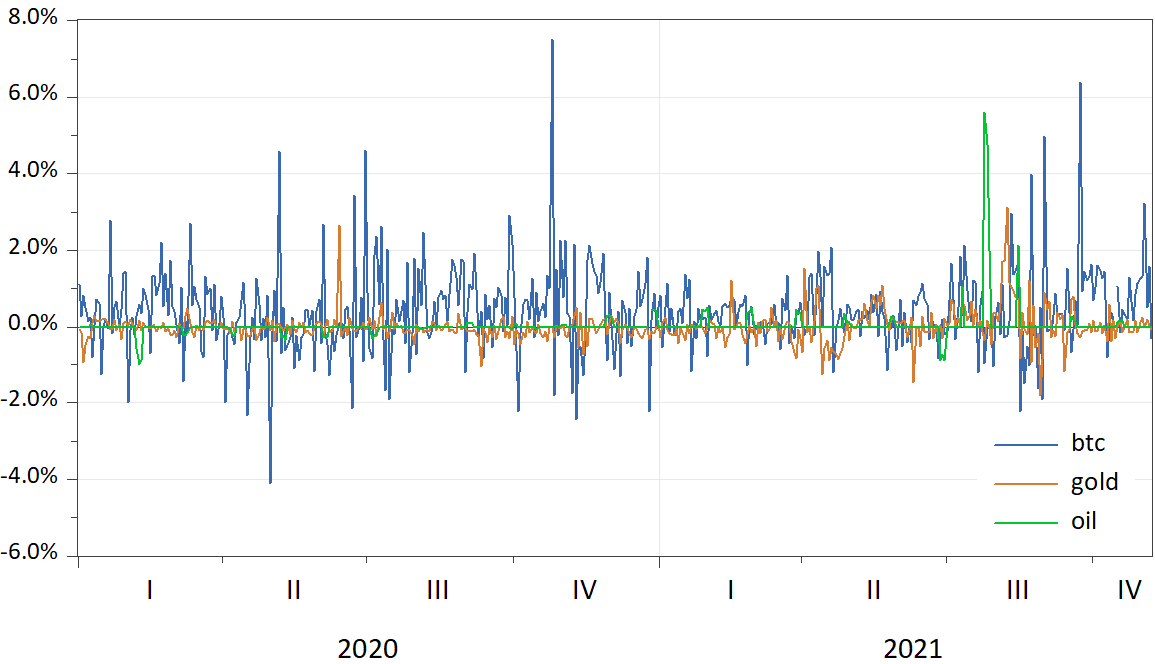}
\caption{Comparison of basis for Bitcoin, gold and oil} \label{fig:basis}
\end{figure}

Table~\ref{tab:descriptive} shows descriptive statistics of the data for the variables used for Bitcoin, gold and oil spot and futures markets in the investigation. 
Looking at the mean, we can see the oil achieves the highest trading volume (0.38 million contracts) and open interest (0.31 million contracts), followed by gold and Bitcoin. Furthermore, key differences are observed in the standard deviation, skewness and kurtosis of the three assets.
The standard deviation for Bitcoin spot and futures prices is larger than those of gold and oil, indicating its volatile nature. While oil spot and futures prices exhibit negative skew to the left. The distribution of open interest is also asymmetric, with a slight left-skew for oil, while it is right-skewed for the other assets.
Regarding kurtosis, the spot price of oil is leptokurtic when compared to the normal distribution\footnote{The kurtosis of a normal distribution is 3.} (with heavier tails), while gold's spot price is platykurtic. In terms of open interest, gold series have heavier tails, while oil series have lighter tails.
Fig.~\ref{fig:basis} displays the basis between January 2020 and November 2021 for the three commodities.

The standard deviations for Bitcoin, gold and oil basis are 1.075, 0.409, 0.386, respectively. The basis is nearly zero-revert, suggesting the overall efficiency \cite{fama_1970} for all assets, while it is relatively weak for Bitcoin, as implied by the larger volatility in Bitcoin basis.


\section{Results}
We tested all possible combinations of maturity ($m$), trading volume ($v$) and open interest ($o$) in explaining the dependent variables, volatility and basis, to estimate Eq. (\ref{eq1-main}). We chose the $mvo$ model for the volatility of all three assets, as it achieved the highest average adjusted $R^2$ ($\bar{R^2}$), consistent with the model investigated in \cite{Ripple2009}. In explaining basis, we chose the $mv$ model for Bitcoin and gold, while the $mvo$ model fits well for oil. 
Table \ref{tab:R1_ModeTables} contains a comparison of the some of the models we tested.
Table \ref{tab:R2_Coeffs} shows the coefficients for the $mvo$ model to explain the volatility of selected contracts (March, June, September and December) for the three assets, however the analysis was conducted on all 47 contracts.


\begin{table*}[tbp]
\centering
\setlength{\tabcolsep}{4pt}

\caption{Model Comparison for the Contract-by-Contract Analysis for Bitcoin, Gold, and Oil Data}\label{tab:R1_ModeTables}
\begin{tabular}{lcrrrrrrlrrrrrrrlrrrrrrrc}
\toprule
    \multicolumn{3}{l}{--- \textit{Volatility} ---}\\    
    & \multicolumn{7}{l}{\bf Bitcoin} &~& \multicolumn{7}{l}{\bf Gold} &~& \multicolumn{7}{l}{\bf Crude Oil}\\
        \cline{1-8}
        \cline{10-16}
        \cline{18-24}
    \addlinespace[0.8ex]
    & \multicolumn{1}{c}{$m$} & \multicolumn{1}{c}{(-)} & \multicolumn{1}{c}{$v$} & \multicolumn{1}{c}{(-)} & \multicolumn{1}{c}{$o$} & \multicolumn{1}{c}{(-)} & \multicolumn{1}{c}{$\bar{R}^{2}$} &  & \multicolumn{1}{c}{$m$} & \multicolumn{1}{c}{(-)} & \multicolumn{1}{c}{$v$} & \multicolumn{1}{c}{(-)} & \multicolumn{1}{c}{$o$} & \multicolumn{1}{c}{(-)} & \multicolumn{1}{c}{$\bar{R}^{2}$} &  & \multicolumn{1}{c}{$m$} & \multicolumn{1}{c}{(-)} & \multicolumn{1}{c}{$v$} & \multicolumn{1}{c}{(-)} & \multicolumn{1}{c}{$o$} & \multicolumn{1}{c}{(-)} & \multicolumn{1}{c}{$\bar{R}^{2}$} \\
        \cline{2-8}
        \cline{10-16}
        \cline{18-24}
\addlinespace[0.8ex]
$m$ & 21.3  & 50.0  & \multicolumn{1}{c}{-} & \multicolumn{1}{c}{-}   & \multicolumn{1}{c}{-}  & \multicolumn{1}{c}{-} & 0.031 &  & 46.8 & 4.5  & \multicolumn{1}{c}{-} & \multicolumn{1}{c}{-}   & \multicolumn{1}{c}{-}  & \multicolumn{1}{c}{-}   & 0.108  &  & 34.0  & 43.8  & \multicolumn{1}{c}{-} & \multicolumn{1}{c}{-}   & \multicolumn{1}{c}{-}  & \multicolumn{1}{c}{-}   & 0.070 \\
$mv$ & \textbf{53.2}   & 0.0  & \textbf{72.3}   & 0.0   & \multicolumn{1}{c}{-}  & \multicolumn{1}{c}{-}   & 0.254 &  & 19.1 & 55.6  & \textbf{53.2}  & 0.0   & \multicolumn{1}{c}{-}  & \multicolumn{1}{c}{-}  & 0.225 &  & 36.2  & 17.6   & 46.8  & 4.5  & \multicolumn{1}{c}{-}  & \multicolumn{1}{c}{-}  & 0.172\\
$mvo$ & 48.9  & 4.3 & \textbf{72.3}  & 0.0  & 19.1 & 100 & 0.280 &  & 12.8 & 0.0  & \textbf{61.7}  & 0.0  & 14.9   & 100  & 0.242  &  & \textbf{55.3} & 3.8  & \textbf{59.6}  & 0.0  & 38.3  & 100  & 0.235 \\
\midrule
\addlinespace[1.0ex]
    \multicolumn{3}{l}{--- \textit{Basis} ---}\\  
    & \multicolumn{7}{l}{\bf Bitcoin} &~& \multicolumn{7}{l}{\bf Gold} &~& \multicolumn{7}{l}{\bf Crude Oil}\\
        \cline{1-8}
        \cline{10-16}
        \cline{18-24}
    \addlinespace[0.8ex]
    & \multicolumn{1}{c}{$m$} & \multicolumn{1}{c}{(-)} & \multicolumn{1}{c}{$v$} & \multicolumn{1}{c}{(-)} & \multicolumn{1}{c}{$o$} & \multicolumn{1}{c}{(-)} & \multicolumn{1}{c}{$\bar{R}^{2}$} &  & \multicolumn{1}{c}{$m$} & \multicolumn{1}{c}{(-)} & \multicolumn{1}{c}{$v$} & \multicolumn{1}{c}{(-)} & \multicolumn{1}{c}{$o$} & \multicolumn{1}{c}{(-)} & \multicolumn{1}{c}{$\bar{R}^{2}$} &  & \multicolumn{1}{c}{$m$} & \multicolumn{1}{c}{(-)} & \multicolumn{1}{c}{$v$} & \multicolumn{1}{c}{(-)} & \multicolumn{1}{c}{$o$} & \multicolumn{1}{c}{(-)} & \multicolumn{1}{c}{$\bar{R}^{2}$} \\
        \cline{2-8}
        \cline{10-16}
        \cline{18-24}
\addlinespace[0.8ex]
$m$ & 48.9 & 13.0 & \multicolumn{1}{c}{-}  & \multicolumn{1}{c}{-}    & \multicolumn{1}{c}{-}   & \multicolumn{1}{c}{-}    & 0.151 &  & 38.3 & 0.0  & \multicolumn{1}{c}{-}  & \multicolumn{1}{c}{-}    & \multicolumn{1}{c}{-}   & \multicolumn{1}{c}{-}    & 0.090 &  & \textbf{87.2} & 43.9 & \multicolumn{1}{c}{-}  & \multicolumn{1}{c}{-}    & \multicolumn{1}{c}{-}   & \multicolumn{1}{c}{-}    & 0.475 \\
$mv$ & \textbf{53.2} & 16.0 & 27.7 & 30.8 & \multicolumn{1}{c}{-}   & \multicolumn{1}{c}{-}    & 0.189 &  & 36.2 & 0.0  & 8.5 & 75.0 & \multicolumn{1}{c}{-}   & \multicolumn{1}{c}{-}    & 0.102 &  & \textbf{85.1} & 45.0 & 36.2 & 41.2 & \multicolumn{1}{c}{-}   & \multicolumn{1}{c}{}    & 0.502 \\
$mvo$ & \textbf{55.3} & 15.4 & 17.0 & 25.0 & 12.8 & 33.3 & 0.205 &  & 27.7 & 23.1 & 17.0 & 75.0 & 12.8 & 16.7 & 0.111 &  & \textbf{87.2} & 48.8 & 10.6 & 100 & 34.0 & 18.8 & 0.554 \\
\bottomrule
\addlinespace[0.8ex]
\multicolumn{24}{l}{The rows denote models and the $m$, $v$ and $o$ columns represent the percentages of contracts for which $m$, $v$, or $o$ are significant out of 47 contracts,}\\
\multicolumn{24}{l}{with values greater than 50\% emphasised in bold. The (-) columns display the percentages of significant contracts with negative coefficients.}
\end{tabular}
\end{table*}

\begin{table*}[tbp]
\centering
\setlength{\tabcolsep}{4pt}

\caption{Regression Parameter Estimates and Related Statistics for Eq. (\ref{eq1-main}) for Selected Contracts - Dependent Variable: Volatility}\label{tab:R2_Coeffs}
\begin{tabular}{cclllrclllrclllrl}
\toprule
    & ~~ &\multicolumn{4}{l}{\bf Bitcoin}~~~ &~& \multicolumn{4}{l}{\bf Gold} &~& \multicolumn{4}{l}{\bf Crude Oil}\\
        \cline{3-6}
        \cline{8-11}
        \cline{13-16}
    \addlinespace[0.8ex]

& \multicolumn{1}{c}{} & \multicolumn{1}{c}{$m$} & \multicolumn{1}{c}{$v$} & \multicolumn{1}{c}{$o$} & \multicolumn{1}{c}{$\bar{R}^{2}$} & \multicolumn{1}{c}{} & \multicolumn{1}{c}{$m$} & \multicolumn{1}{c}{$v$} & \multicolumn{1}{c}{$o$} & \multicolumn{1}{c}{$\bar{R}^{2}$} & \multicolumn{1}{c}{} & \multicolumn{1}{c}{$m$} & \multicolumn{1}{c}{$v$} & \multicolumn{1}{c}{$o$} & \multicolumn{1}{c}{$\bar{R}^{2}$} \\
\midrule

Mar-18& &  2.091** &  0.019 &  -0.008 &  0.153  &  & -0.020 & 4.103* & 4.395   & 0.231   &  & 0.043**& 0.072***  & -0.083***  & 0.621  & \\
& &  (0.031) &  (0.156)  &  (0.674)   &     &   & (0.146)& (0.077)& (0.125) & &   & (0.037)& (0.000)& (0.000) & & \\
Jun-18& &  0.199*  &  0.004*** &  0.000 &  0.187  &  & -0.002 & 0.009**& -0.001  & 0.120   & & 0.063***  & 0.033**& -0.029  & 0.171   &\\
& &  (0.058) &  (0.003)  &  (1.000)   &     &   & (0.741)& (0.036)& (0.875) & &   & (0.007)& (0.019)& (0.218) & & \\
Sep-19& &  0.181*  &  0.003**  &  -0.002 &  0.081  &   & -0.019*& 4.382**& 2.447** & 0.245   & & 0.072**& 0.061* & -0.060  & 0.069   & \\
& &  (0.087) &  (0.018)  &  (0.443)   &     &   & (0.058)& (0.020)& (0.081) & &   & (0.024)& (0.053)& (0.158) & & \\
Dec-18& &  0.007 &  0.013*** &  -0.010 &  0.160  &   & 0.005  & 0.014**& -0.009  & 0.241   &   & 0.047  & 0.098***  & -0.149**& 0.349   & \\
& &  (0.988) &  (0.010)  &  (0.173)   &     &   & (0.369)& (0.019)& (0.213) & &   & (0.299)& (0.001)& (0.005) & & \\
Mar-19& &  0.116   &  0.001 &  -0.001 &  -0.028 &   & -0.002 & 1.339  & 0.527   & 0.074   &   & 0.305***  & 0.068  & -0.200* & 0.262   & \\
& &  (0.226) &  (0.336)  &  (0.613)   & &   & (0.663)& (0.129)& (0.655) & &   & (0.005)& (0.321)& (0.092) & & \\
Jun-19& &  -0.129 &  0.006*** &  -0.022*** & 0.512  &   & 0.007  & 0.003* & -0.013  & -0.013  &   & 0.050**& 0.064***  & -0.063**& 0.323   & \\
& &  (0.777) &  (0.000)  &  (0.001)   & &   & (0.235)& (0.575)& (0.135) & &   & (0.039)& (0.000)& (0.041) & & \\
Sep-19& &  0.395** &  0.011*** &  -0.011*** & 0.642  &   & -0.012 & 6.066***  & -2.415  & 0.455   &   & 0.103* & 0.142***  & -0.169**& 0.266   & \\
& &  0.018   &  (0.000)  &  (0.000)   & &   & (0.450)& (0.000)& (0.181) & &   & (0.069)& (0.001)& (0.031) & & \\
: &  & \multicolumn{1}{c}{:} && \multicolumn{1}{c}{:}  &   && \multicolumn{1}{c}{:} & & \multicolumn{1}{c}{:}  &   && \multicolumn{1}{c}{:} & & \multicolumn{1}{c}{:}  & \\
Jun-21& &  0.442   & 0.010& -0.015**  & 0.039  &   & -0.002 & 0.012  & -0.006  & -0.054  &   & 0.202***  & 0.112**& -0.263***  & 0.413   & \\
& & (0.644) & (0.073)  & (0.075)   & &   & (0.909)& (0.382)& (0.774) & &   & (0.000)& (0.046)& (0.000) & & \\
Sep-21& &  0.668***& 0.010& -0.006*** & 0.518  &   & 0.002  & 8.047***  & -2.194  & 0.165   &   & 0.101**& 0.235***  & -0.151**& 0.325   & \\
& & (0.003) & (0.000)  & (0.004)   & &   & (0.894)& (0.006)& (0.617) & &   & (0.050)& (0.000)& (0.034) & & \\


\bottomrule
\addlinespace[0.8ex]
\multicolumn{15}{l}{\textsuperscript{***}, 
  \textsuperscript{**}, 
  \textsuperscript{*} : statistically significant at the 1\%, 5\%, and 10\% levels, respectively.
  P-values are displayed in parentheses.}
\end{tabular}
\end{table*}

\subsection{Volatility}
The $mvo$ model was chosen to explain volatility upon examining Table \ref{tab:R1_ModeTables}. Aside from having the highest average $\bar{R^2}$, $m$ and $v$ are generally significant for all three assets in this model. As for $o$, while not significant for more than 20\% of the contracts for Bitcoin and gold, it is relatively significant (38.3\% of contracts) for oil and always has a negative coefficient when significant for all three assets - indicating underlying structure. The importance of $v$ is emphasised in the $mv$ model which always has $v$ as the most significant explanatory variable. $v$ is always positive, indicating that volatility increases with the daily trades, as expected.

As $o$ always has a negative effect on all 3 assets, it appears to offset the positive influence of $v$ on volatility, consistent with \cite{Ripple2009}. One explanation for this is that higher $o$ denotes that more contracts are available to be traded - representing increased market depth, implying greater liquidity. However, it is interesting to note that increased $v$ traditionally also implies greater liquidity, yet $o$ and $v$ have opposite effects on volatility. This could be because $o$ represents the number of contracts that have not yet been exercised, 
implying the increase in the market depth \cite{Ripple2009} and so have the effect of stabilising price. Meanwhile, $v$ represents completed trades whose effect on volatility has already been priced in. Unlike \cite{Ripple2009}, our results do not show that $o$ only offsets half of the positive effect of $v$ on average. 


$m$ is positive for the majority of contracts across the 3 assets, negative for at most 4.3\% of contracts. We propose two possible explanations for this, the first being that there is less volume, $v$, far from maturity which leads to higher volatility as the contract is being traded less. Alternatively, we can use $o$ as a measure of market participation and note that $o$ starts at 0 and grows slowly far from maturity, as seen in Fig. \ref{fig:BTC_oi}, implying that the market is more prone to manipulation due to low market participation, resulting in increased volatility.

\subsection{Basis}

Upon examining Table \ref{tab:R1_ModeTables}, the $mvo$ model was chosen to explain basis for oil as it yields the highest average $\bar{R^2}$ of 0.554 and aside from $v$, all explanatory variables appear to be significant. The $mv$ model also yields a high average $\bar{R^2}$ of 0.502, with a more significant role played by $v$, suggesting that the model could be fruitful for future research, however previous research \cite{Ripple2009} suggests that $o$ is an important explanatory variable for oil. The $mv$ model was chosen for Bitcoin and gold as it yields a similar $\bar{R^2}$ to the $mvo$ model, however $o$ is not significant for more than 6 contracts for these assets and it is favourable to use the minimum number of explanatory variables required to describe the basis.

In contrast with volatility, when significant, $o$ has a generally positive impact on basis, at a minimum of 66.7\% of contracts for Bitcoin and a maximum of 83.3\% of contracts for gold. A possible explanation for this is that a large $o$ represents a large number of unsettled futures contracts whose impact on the futures price has not yet been priced in. Meanwhile, such an effect does not exist in the spot market, allowing a gap between the prices of the two markets to open as this manifests as a lag between them. 

$v$ generally has a negative effect on basis at 100\% and 75.0\% of significant contracts for oil and gold respectively. This is unsurprising as spot trading volume is expected to be correlated to futures trading volume, so we expect the volatility of both markets to decrease with increasing $v$, rendering it easier for the spot and futures prices to track each other. However, $v$ is only negative for 25\% of significant contracts for Bitcoin, suggesting that there is less correlation between spot and futures trading volume for the asset. This could be because the market for Bitcoin is less mature than the other two markets, despite recent development in the market for Bitcoin \cite{AkyildirimETH2020}, and perhaps because most market participants predominantly trade spot Bitcoin. 

$m$ is mostly positive when significant for Bitcoin and is always positive for gold in the $mv$ model, indicating that far from maturity, there is a larger discrepancy between spot and futures prices due to low market participation in the futures contract. This is consistent with the largely positive impact of $o$ in the $mvo$ model for these assets, as $o$ starts from 0 far from maturity, as previously noted.

\section{Conclusion}

In this paper, we applied contract-by-contract analysis, as previously employed in crude oil futures volatility investigations, for spot and futures daily prices. We used maturity, trading volume and open interest data to investigate the determinants of the volatility of futures prices for three commodities: gold, oil and Bitcoin. This methodology was then extended to investigate the determinants of basis.
By using data from 18th December 2017 to 30th November 2021, we found that maturity affects volatility positively for Bitcoin and oil, indicating that less market participation when far from maturity is one of the main causes of the volatility for these two assets. We also certified a positive and significant role for trading volume in determining the volatility for all three assets, indicating that volatility rises as the number of daily trades increases, as expected. Furthermore, we found a possible negative influence (when significant) of open interest in shaping the volatility in these three assets. 
Additionally, we verified that maturity affects the basis of Bitcoin and gold positively - confirming the general theory that the basis converges to zero at maturity, but not for oil.



%
%
\bibliographystyle{splncs04}
\bibliography{bibliography}
%

\end{document}